\documentclass[pra,aps,%
 reprint,%
 twocolumn,
superscriptaddress,
nofootinbib,
longbibliography]{revtex4-2}
\usepackage{makecell} 
\usepackage{braket}
\usepackage[table,dvipsnames]{xcolor}

\usepackage{graphicx}
\usepackage{amsfonts}
\usepackage{amssymb}
\usepackage{booktabs}
\usepackage{amsmath}
\usepackage{amsthm}
\usepackage[colorlinks=true,linkcolor=teal,citecolor=teal,urlcolor=teal,plainpages=false,pdfpagelabels]{hyperref}
\usepackage{color}

\usepackage{mathtools}
\usepackage{bbm}
\usepackage{textcomp}

\usepackage{subcaption}

\usepackage{array}

\usepackage{enumitem}
\usepackage{verbatim}

\usepackage[utf8]{inputenc}
\usepackage[T1]{fontenc}
\usepackage{etoolbox}

\usepackage{newtxtext}
\usepackage{newtxmath}
\usepackage[normalem]{ulem}

\usepackage[utf8]{inputenc}
\usepackage[T1]{fontenc}
\usepackage{ragged2e}

\makeatletter
\g@addto@macro\bfseries{\boldmath}
\makeatother

\theoremstyle{definition}

\renewcommand{\qedsymbol}{$\blacksquare$}

\renewcommand{\qedsymbol}{\unskip\nobreak\quad\qedsymbol}
\renewcommand{\qedsymbol}{$\blacksquare$}

\definecolor{dblue}{rgb}{0.0, 0.53, 0.74}

\begin{document}

\title{Fictitious Copy Quantum Error Mitigation}

\author{Akib Karim$^*$} \affiliation{Data 61, CSIRO, Research Way Clayton 3168, Victoria, Australia} 
\author{Harish J. Vallury$^*$}\affiliation{Data 61, CSIRO, Research Way Clayton 3168, Victoria, Australia}
\author{Muhammad Usman} \affiliation{Data 61, CSIRO, Research Way Clayton 3168, Victoria, Australia}\affiliation{School of Physics, The University of Melbourne, Parkville 3010, Victoria, Australia} 

\begin{abstract}

\noindent
Errors are arguably the most pressing challenge impeding practical applications of quantum computers, which has instigated vigorous research on the development of quantum error mitigation (QEM) techniques. Existing QEM methods suppress errors with a varying degree of efficacy but importantly demand significant additional quantum and classical computational resources. In this work, we present Fictitious Copy Quantum Error Mitigation (FCQEM) method which corrects quantum errors without requiring any additional quantum resources and purely relies on using classical postprocessing of a joint probability distribution to correct expectation values. The joint probability distribution can be measured ``fictitiously'' by sampling one copy of noisy quantum circuit twice, or classically squaring probabilities from simply one copy. We show that FCQEM can recover eigenvalues even if exact eigenstates are not prepared. Furthermore, our technique can benefit other noise mitigation techniques with no additional quantum resources, which is demonstrated by combining FCQEM with the Quantum Computed Moments (QCM) method. FCQEM can compensate for noise that is pathological to QCM, and QCM allows for FCQEM to recover the ground state energy with a larger variety of trial states. We show that our technique can find the exact ground state energy of molecular and spin models under simulated noise models as well as experiments on a Rigetti 82-qubit  superconducting quantum processor. The reported FCQEM method is general purpose for the current generation of quantum devices and is applicable to any problem that measures eigenvalues of operators on sharply peaked distributions.
 
\end{abstract}

\maketitle
\def\thefootnote{*}\footnotetext{These authors contributed equally to this work}\def\thefootnote{\arabic{footnote}}

\section{Introduction}

Quantum computers are anticipated to run algorithms with better scaling of computational resource requirements with problem size than classical computers, thereby allowing calculations that are not currently possible within any reasonable timescales on classical computing platforms~\cite{Montanaro2016,Shor1994}. In other instances such as in the case of reliability issues of machine learning models, quantum computing may offer robustness, enabling trustworthy applications \cite{West2023}. Despite significant theoretical developments underscoring the promise of quantum computing's computational advantage, its implementation in practical applications is still an ongoing endeavor. A major challenge hindering the transformation of theoretical ideas into experimental realization is limitations imposed by the noise or errors in the current generation of quantum computers that destroy quantum correlations and reduces the information in algorithmic outputs. This severely restricts both the types of algorithms and the scale of the algorithms that can be executed on a quantum processor. There is significant research into enabling fault tolerant quantum computers in the future~\cite{Shor1996,Gicev2023,Kang2026} but the development of quantum processors commensurate with the fault-tolerance resource requirements may take many more years. Meanwhile, a highly active area of research focuses on error mitigation strategies (also known as noise mitigation) \cite{Nachman2020, Endo2018, Temme2017, Huang2020, Akib2018, Huggins2021, Koczor2021, Akib2024} to enable applications even for the current generation of small-scale quantum devices.  

Typically, quantum error mitigation strategies use additional quantum resources to suppress errors and obtain better algorithm performance. This can be in terms of extra measurements on small characterizing circuits~\cite{Nachman2020, Endo2018}, or variations on the specific circuit for the algorithm~\cite{Temme2017, Huang2020}. In particular, additional qubits to duplicate the algorithm and using entangled projections between copies can increase the information capacity with finite uses of the channel~\cite{Akib2018}. This was the principle behind Virtual Distillation (VD)~\cite{Huggins2021,Kozcor2021Implement}, which uses duplicate circuits on additional qubits to measure expectation values on powers of the density matrix and amplify the dominant eigenvector. However, the original circuit implementations either only worked with 1-local Hamiltonians~\cite{Huggins2021} or required an ancilla with a deep controlled derangement operator~\cite{Kozcor2021Implement} with the exact implementation requiring very deep circuits. For practical calculations, the density matrix was often approximated using classical shadow tomography~\cite{Alireza2023} and powers taken in postprocessing, which needs exponential measurements for arbitrary states. It was found that the exact implementation can be decomposed into a series of shallow circuits~\cite{Akib2024}, however this required measuring every circuit without any truncation possible. The development of an effective QEM technique with minimal or no resource requirements is still an open question. 

In this work, we present Fictitious Copy Quantum Error Mitigation (FCQEM) method which does not require any additional quantum resources but relies only on classical post-processing. The idea of the FCQEM technique stems from showing that VD admits a series expansion that can be truncated to first order and then approximated classically. We demonstrate that truncated VD recovers exact expectation values when measuring eigenstates of a Hamiltonian and, in fact, expectation values can tend towards the nearest eigenstate. This means we can measure eigenvalues of operators without preparing the exact eigenvectors as a trial state. We also discover an additional classical approximation to the truncated operator assuming the two duplicates are identical. By using these ideas, we formulate the FCQEM method, which takes powers of the sampled probability distribution rather than powers of the density matrix and requires no additional quantum resources, in particular, does not require preparing and entangling two copies of the entire circuit as per VD. We show that when estimating operators by decomposing into measurement bases, if the distributions are sharply peaked then FCQEM performs well with the limit of the eigenbasis where FCQEM is identical to truncated VD. This makes the method particularly effective for finding eigenvalues of diagonally dominant Hamiltonians when decomposing into Pauli strings that include the computational basis, e.g., Full Configuration Interaction and Ising model Hamiltonians.

We quantify the state-dependence of the truncation error between FCQEM and full VD and note that, for many problems, FCQEM alone may not perform ideally, e.g., when finding the ground state energy of a Hamiltonian where the ground state itself does not show a sharply-peaked distribution in the Hamiltonian measurement bases. However, FCQEM requires only straightforward classical post-processing and no additional quantum overhead, so we may easily incorporate it with other techniques that help in these scenarios. In particular, in this work we combine FCQEM with the quantum computed moments (QCM) method for ground state energy estimation~\cite{hv2020, Jones2022, hv2023_2}, whereby expectation values of powers of the Hamiltonian are calculated with respect to an inexact trial state to produce the true ground state energy. Crucially, the robustness of QCM to suboptimal trial states allows us to choose regimes that mitigate the state-dependent truncation error of FCQEM relative to full VD. Remarkably we find that FCQEM improves the QCM ground state energy estimate in noisy situations where QCM does not perform optimally. By performing experimental implementations of 4-qubit HeH$^+$ molecular and 10-qubit spin chain Hamiltonians on real superconducting devices from Rigetti, we conclusively demonstrate the combined efficacy of FCQEM and QCM methods by fully recovering ground state energies within $10^{-5}$ Ha of the exact value for HeH$^+$ and $0.3\%$ for the 10-qubit Ising spin chain. Finally, we note the scalability of this approach as it can be easily incorporated into existing schemes with no additional quantum overhead. We study its effectiveness under noisy simulation of the larger water molecule and spin models in the regime of hundreds of qubits and show it is able to recover the ground state energy.

\begin{figure*}
    \centering
    \includegraphics[width=0.8\linewidth]{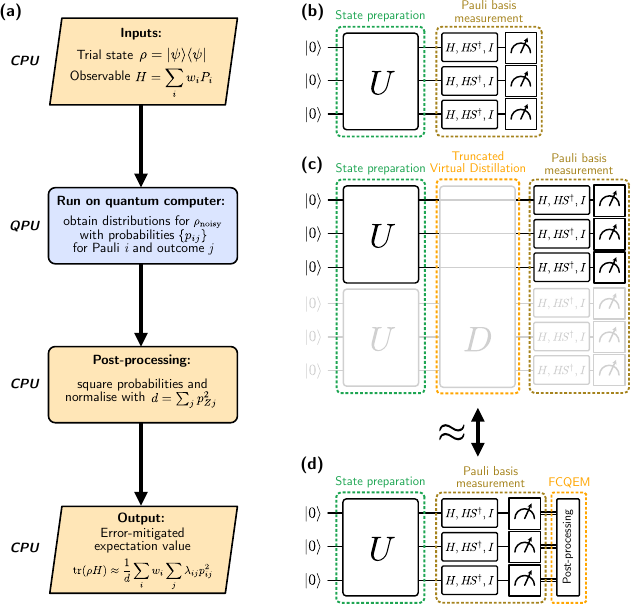}
    \caption{\justifying \textbf{FCQEM Framework.} (a) Flowchart outlining the Fictitious Copy Quantum Error Mitigation (FCQEM) technique, along with circuit-level descriptions depicted in (b-d). An $n$-qubit state is prepared on a quantum processor and measured in Pauli basis as illustrated in (b), however the noise in quantum processor would severely degrade the state fidelity. In (c) we show a circuit diagram relating FCQEM to a truncated Virtual Distillation approach, but requiring no quantum overhead which is equivalent to canceling the noise through classical post-processing step in (d). As will be demonstrated in our work for a variety of implementations on a Rigetti quantum processor, FCQEM recovers the trial state from the noisy distribution in the computational basis.}
    \label{fig:flowchart}
\end{figure*}

\section{Theory}
In this section, we derive the theory for the FCQEM technique. We start from Virtual Distillation (VD)~\cite{Huggins2021,Kozcor2021Implement}, which requires prohibitively deep entangling circuits over multiple copies. We have previously shown in Karim \textit{et al}.~\cite{Akib2024} how to use a tailored circuit decomposition to write the circuits as a series of low depth circuits, however these required all terms in the series to be measured. Here we use a power series to expand the entangling circuits and truncate to first order. We then derive a classical post-processing method with no quantum resources that does not even require preparing multiple copies as it uses "fictitious" copies. Thus we avoid the deep entangling circuits of virtual distillation and the large number of measurements from circuit decomposition, which allows its use on near term hardware. The proposed full method for FCQEM is given as Figure~\ref{fig:flowchart}.

\subsection{Fictitious Copy Quantum Error Mitigation}

In the original proposals, Virtual Distillation~\cite{Huggins2021, Kozcor2021Implement} corrects observables $M$ on arbitrary quantum states $\rho$ by measuring them on $\rho^2$. Explicitly,

\begin{equation}
    tr(M\rho)\approx\frac{tr(M\rho ^2)}{tr(\rho^2)}  = \frac{tr(M_2 S\rho\otimes\rho)}{tr(S\rho\otimes\rho)} ,
\end{equation}

\noindent
where $S$ is the operator that flips between the two subsystems, i.e. acting on an arbitrary state $\sum c_{i,j}\ket{i,j}$:
\begin{equation}
    S\sum c_{i,j}\ket{i,j} = \sum c_{j,i}\ket{i,j},
\end{equation}

\noindent
and $M_2$ is the symmetrised observable:
\begin{equation}
    M_2 = \frac{1}{2}(M\otimes I + I\otimes M).
\end{equation}

\noindent
Since $S$ is unitary, we can decompose it as:
\begin{equation}
    S = e^{-iG} D e^{iG},
\end{equation}
where $G$ is Hermitian and $D$ is a diagonal. Using the power series expansions, we can write this as:

\begin{equation}
    \left( I + \frac{-iG}{1!} + \frac{(-iG)^2}{2!} + \ldots \right) D \left( I + \frac{iG}{1!} + \frac{(iG)^2}{2!} + \ldots \right).
\end{equation}

\noindent
Expanding gives us:

\begin{equation}
    D + \left( -\frac{iG}{1!} D + D \frac{iG}{1!} \right) + \ldots
\end{equation}

\noindent
This results in a series of terms that can be grouped based on the powers of $ G $.

\begin{equation}
    S = D + i[G,D] -\frac{1}{2}[G,[G,D]]  + \ldots \label{eq:series}
\end{equation}

\noindent
To first order, the full virtual distillation correction can be approximated by:
\begin{equation}
\frac{tr(M S\rho\otimes\rho)}{tr(S\rho\otimes\rho)} \approx
\frac{tr(M D\rho\otimes\rho) }{tr(D\rho\otimes\rho) }\label{eq:VVD} 
\end{equation}

\noindent
This $D$ is a diagonal matrix in the computational basis such that:
\begin{equation}
    D\sum c_{i,j}\ket{i,j} = \sum
    \begin{cases}
        c_{i,j}\ket{i,j}& \text{if } i\le j\\
        -c_{i,j}\ket{i,j}& \text{if } i > j\\
    \end{cases}
\end{equation}
where the inequalities can equivalently be defined in reverse. $D$ preserves the coefficients when $\ket{i,j} = \ket{j,i}$. When $i\ne j$, it gives $+1$ to the corresponding symmetric combination in $S$ and $-1$ to the antisymmetric combination. Since $D$ is a diagonal matrix in the chosen basis, we no longer require entangling gates on the duplicate copies of the state when measuring in that basis. We can therefore derive a classical post-processing technique. 

For each measurement operator, $M$, we measure $U\rho U^{-1}$, where $U^{-1}MU$ is diagonal. This is measured in the computational basis to give a discrete probability distribution over all possible qubit outcomes. Let $X$ be a discrete random variable over such outcomes where:
\begin{equation}
    P(X=i) = p_i.
\end{equation}

\noindent
A measurement of $M$ can be performed by first decomposing $M$ into eigenvalues, $M = \sum_i \lambda_i\ket{i}$. The expectation value of $X$ is now given by:
\begin{equation}
    E_X[M] = \sum_i \lambda_i p_i.
\end{equation}

\noindent
We construct a joint probability mass function over two copies of $X$ as $p_{X_1,X_2}(i,j)$, which we simplify to $P_{ij}$ for ease of notation defined as:

\begin{equation}
    P_{ij} =  P(X1=i, X2=j) = p_ip'_j.
\end{equation}

\noindent
The effect of measuring the symmetrised operator $M_2 = \frac{1}{2}(M\otimes I + I\otimes M)$ is 
\begin{equation}
    E_{X\otimes X}[M_2] = \frac{1}{2}\left( \sum_i \sum_j\lambda_iP_{ij}  + \sum_j\sum_i\lambda_jP_{ij} \right). 
\end{equation}

\noindent
We can pull the $\lambda$ out of the interior sums to give:
\begin{equation}
\frac{1}{2}\left( \sum_i \lambda_i\sum_jP_{ij}  + \sum_j\lambda_j\sum_iP_{ij} \right).
\end{equation}

\noindent
We now look at the effect of multiplying by $D$. In this formalism, this sets:
\begin{equation}
D[P_{ij}] \to
\begin{cases}
-P_{ij}, & \text{if } i<j,\\
P_{ij}, & \text{if } i\ge j.
\end{cases}
\end{equation}

\noindent
The effect of measuring $M_2$ on this object is:
\begin{equation}
\begin{split}
E_{D}[M_2] =\frac{1}{2} \sum_i^n \lambda_i\left(\sum_{j=0}^{i}P_{ij} - \sum_{j=i+1}^nP_{ij} \right) + \\
\frac{1}{2}\sum_j^n\lambda_j\left(\sum_{i=j}^nP_{ij}-\sum_{i=0}^{j-1}P_{ij} \right),
\end{split}
\end{equation}
where we have explicitly summed to the total number of outcomes $n$, which is $2^q$, where $q$ is the number of qubits. Now, we note that $P_{ij} = p_ip'_j$ to further pull out factors from the interior sums, relabel and simplify to give:
\begin{equation}
\frac{1}{2} \sum_i^n \lambda_i\left((2p_ip'_i)+\sum_{j=0}^{i-1}(p_ip'_j-p_i'p_j) + \sum_{j=i+1}^n(p'_ip_j-p_ip'_j) \right).~\label{eq:FCQEM2}
\end{equation}

\noindent
We can sample two different $X$ from the same distribution to see different noise models. However, we consider only recovering one $X$ and duplicating it exactly to form the joint mass function, i.e. $p' = p$. In this case all the terms cancel except the squared term. The expectation value is therefore:
\begin{equation}
 tr(MD\rho \otimes\rho) =\sum_i^n \lambda_ip_i^2\label{eq:FCQEM}.
\end{equation}

Similarly, 
\begin{equation}
 tr(D\rho \otimes\rho) =\sum_i^n p_i^2.
\end{equation}

This analysis assumes we can measure $M$ in an eigenbasis, however typically we decompose $M$ into Paulis or some other projective measurement as $M = \sum_i w_i\hat{Q_i}$ for weights $w$ over projectors $\hat{Q}$. In this case, we obtain a probability distribution for each projector, which we can combine into the expectation value using the weights:
\begin{equation}
    tr(MD\rho \otimes \rho) = \sum_i w_i \sum_j \lambda_{ij}p_{ij}.
\end{equation}

Furthermore, there is a choice of which basis $D$ is diagonal in. We denote this basis with $i=Z$ and provide discussion on the implications of the choice in the following section.

Our final expression for the corrected expectation value is:
\begin{equation}
    tr(M\rho) \approx \frac{\sum_i w_i \sum_j \lambda_{ij}p_{ij}}{\sum_j \lambda_{Zj}p_{Zj}},~\label{eq:FCQEM_full}
\end{equation}

for using one set of sampling statistics possibly from existing data. If multiple sampling statistics are available, we can create a separate fictitious copy using Equation~\ref{eq:FCQEM2}.

Therefore, where Virtual Distillation relied on measurements on the powers of the density matrix, Fictitious Copy Quantum Error Mitigation uses powers of the measured probability distribution normalised over a preferred basis. 

\subsection{Conditions for Equivalence with Virtual Distillation}

Normally, to measure $tr(MS\rho)$, we find $P$ that diagonalises $M_2S$, i.e. $D^* = P^{-1} M_2S P$, which is possible since we construct $M_2$ such that $[M_2,S] = 0$. We then measure:
\begin{equation}
    tr(M_2 S\rho) = tr(PD^*P^{-1}\rho) = tr(D^*P^{-1}\rho P).
\end{equation}
We can then find a circuit that rotates $\rho$ by $P^{-1}$ and measure in the computational basis. 

This approximation will not recover the exact observables for all possible states and the truncation error is state dependent of order $O([G,D]\rho)$. However, we show that the approximation is exact for eigenstates. Consider Equation~\ref{eq:VVD}.

\begin{equation}
    \frac{tr(SM\rho\otimes\rho)}{tr(S\rho\otimes\rho)} = \lambda \left( \frac{tr(S\rho\otimes\rho)}{tr(S\rho\otimes\rho)} \right) = \lambda,
\end{equation}

\begin{equation}
    \frac{tr(DM\rho\otimes\rho)}{tr(D\rho\otimes\rho)} = \lambda \left( \frac{tr(D\rho\otimes\rho)}{tr(D\rho\otimes\rho)} \right) = \lambda,
\end{equation}

\noindent
i.e. both RHS and LHS cancel to $\lambda$ the correct eigenvalue. For states that are not eigenvectors, it is not guaranteed that the noiseless case will recover the correct value due to the truncation error in the series given in Equation~\ref{eq:series}. We can evaluate the derivative given a small rotation to identify how the value will change if a state that is not the eigenstate is prepared. Define the state rotated away from the eigenvector to be: $\rho'(t) = e^{iHt} (\rho \otimes \rho)~ e^{-iHt}$, where without loss of generalisation, we pick an arbitrary Hermitian matrix $H$ that we rotate with angle $t$. Our corrected measurement is:
\begin{equation}
    f(t) = \frac{tr(DM\rho')}{tr(D\rho')},
\end{equation}

\noindent
where we note that $f(0) = \lambda$. We calculate the derivative with respect to $t$:
\begin{equation}
    f'(t)=i\frac{tr([H,DM]\rho')tr(D\rho') - tr(DM\rho')tr([H,D] \rho')}{tr(D\rho')^2}.
\end{equation}
We evaluate this at $t=0$ where $\rho' = \rho\otimes\rho$ and note that $M\rho = \lambda \rho$ to give:
\begin{equation}
    f'(0) = i\frac{tr(D(\lambda I-M)H\rho\otimes\rho)}{tr(D\rho\otimes\rho)}.
\end{equation}
We can swap $M$ and $H$ using their commutator to cancel out the eigenvalue term. We have the final expression for when preparing incorrect states in the $H$ direction will still recover the eigenvalue:
\begin{equation}
    f'(0) = i\frac{tr(D[H,M]\rho\otimes\rho)}{tr(D\rho\otimes\rho)} = 0.~\label{eq:err_deriv}
\end{equation}
We can thus see that when the prepared state, $\rho$, is close to the eigenstate in a direction $H$ such that $[H,M]=0$; or when the modified distribution $D\rho$ is orthogonal to $[H,M]$, then we will measure the eigenvalue exactly. This means we are free to choose a trial state that is not the eigenstate and still recover the eigenvalue. It also means we can pick multiple $M$ and find their eigenvalues using the same data from measuring $\rho$. This will significantly reduce the quantum resources used for algorithms that require sweeping a family of Hamiltonians such as finding potential energy surfaces for quantum chemistry or phase diagrams for spin models.

Finally, we consider the case of decomposing the measurement $M$ into Pauli strings. We have:
\begin{equation}
    M = \sum w_i\hat{Q_i},\label{eq:paulidecomp}
\end{equation}

\noindent
where we use local rotations to diagonalise each $Q_i$ and take $tr(M\rho) = \sum w_itr(Q_i \rho)$. For our full calculation we have:
\begin{equation}
    tr(M_2 D\rho^{\otimes2} ) = \sum w_itr(Q_i D\rho^{\otimes2}),
\end{equation}
however, now when we find $P_i$ to diagonalise each $Q_i$, 
\begin{equation}
    tr(Q_i DP_i^{-1}\rho^{\otimes2}P_i) = tr(P_iQ_i DP_i^{-1}\rho^{\otimes2}),
\end{equation}
we need this to equal $tr(P_iQ_i P_i^{-1}D\rho^{\otimes2})$ so that we can simplify the calculation to use squares of probabilities as per section B. However, this is only true up to error term:

\begin{equation}
    \epsilon = w_itr(P_i Q_i[D,P_i^{-1}]\rho^{\otimes 2}).~\label{eq:error}
\end{equation}

Since we have chosen $D$ to be the diagonal first order truncation of $S$, it requires us to pick a basis for it to be diagonal in, i.e. when we picked $j=Z$ in Equation~\ref{eq:FCQEM_full}. In the ideal case, we would pick the eigenbasis of the operator, $M$ and use projectors with no additional error, however we can pick a basis that is close to the operator eigenbasis to minimise the error. For instance, we can pick the computational basis, which means $D$ will commute with the diagonalising operators of all Pauli strings with $Z$ and $I$, i.e. $Q_i = \{Z,I\}^{\otimes n}$. In particular, this works only for diagonally dominant Hamiltonians, specifically ones where the weights $w_i$ are larger for strings of the form $Q_i = \{Z,I\}^{\otimes n}$. Fortunately, this is the case for a large variety of physically relevant Hamiltonians such as for Full Configuration Interaction or Ising models. 

\section{Methodology and Results}
\subsection{Numerical Simulation}
To evaluate FCQEM numerically, we calculated the Full Configuration Interaction Hamiltonian for HeH$^+$ with a STO-3G basis set~\cite{Pople2003} using PySCF~\cite{Sun2017}. We performed Hartree-Fock at a distance of $1$Å  for the positive charge state to get the molecular orbital basis and Hartree-Fock state.

To analyse the effect of incoherent noise and compare against VD, the exact state was found by direct diagonalisation and a global depolarising noise channel was applied to the density matrix:
\begin{equation}
    \rho \rightarrow (1-p)\rho + p\frac{I}{d},
\end{equation}
where $p$ controls the strength of the depolarisation channel, $I$ is the identity and $d$ is the dimension of the density matrix.

Similarly, to analyse coherent noise, the unitary for a rotation around the Y-axis was multiplied to the last qubit. Additional single qubit basis gates were applied to measure along each Pauli basis and the probability for each outcome calculated from the density matrix. The probability distribution was then corrected with FCQEM as per Equation~\ref{eq:FCQEM_full}. This is compared against VD to determine the truncation error.

\begin{figure}

    \flushleft{(a)}
    \includegraphics{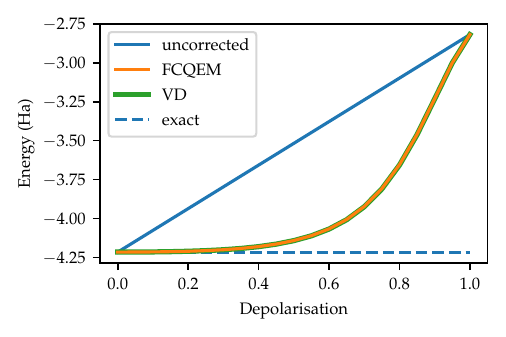}
    \vspace{-1.1cm}
    \flushleft{(b)}
    \includegraphics[]{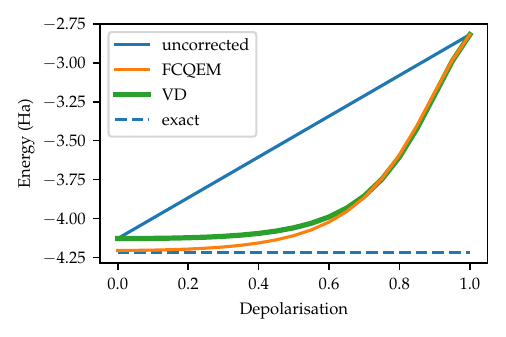}
    \vspace{-1.1cm}
    \flushleft{(c)}
    \includegraphics[]{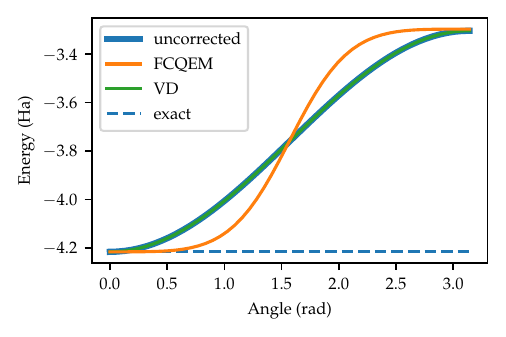}
    
    \caption{\justifying \textbf{Depolarisation Noise Simulations.} Comparison of Virtual Distillation and Fictitious Copy Quantum Error Mitigation on HeH$^+$ under a totally depolarising noise channel before measurement over all four qubits. (a) shows exact state preparation before noise channel; (b) shows applying a Y rotation of $0.2\pi$ on the fourth qubit which prepares the incorrect ground state; (c) shows varying the Y-rotation from the exact ground state at $0$ to $\pi$. Exact refers to exact diagonalisation of the Hamiltonian for the ground state energy. There is no truncation error for exact eigenstates, shown here with the ground state. Furthermore, FCQEM flattens the potential energy surface around this state, allowing for inexact states to correct towards the ground state}
    \label{fig:H2}
\end{figure}

The results are shown in Figure~\ref{fig:H2}. We first compare between VD and FCQEM for the exact ground state of HeH$^+$ molecule given in Figure~\ref{fig:H2}(a). FCQEM has no truncation error for zero noise and closely matches VD as depolarisation strength increases. In this case, FCQEM is able to correct very closely to VD, however in general, FCQEM will not be exactly the same since noise may add contributions to the density matrix that are in a direction such that Equation~\ref{eq:err_deriv} is non-zero and so will introduce a small error.

We now prepare a state that is not an eigenstate, which is shown in Figure~\ref{fig:H2}(b). At no noise, FCQEM is unable to recover the observables due to the VD truncation error. However, it gives energy values closer to the ground state energy as seen in Equation~\ref{eq:err_deriv}. To investigate this, in Figure~\ref{fig:H2}(c), we sweep the angle around the $Y$ axis of the first qubit at no depolarising error. This effectively shows us the truncation error as a function of state. There is no truncation error for the eigenstate at $0$ angle and small perturbations around the ground state do correct to the ground state energy. However, for angles between $\frac{\pi}{2}$ and $\pi$, the expectation values tend towards the value at $\pi$, which leads to maximum truncation error at $\frac{3\pi}{4}$. 

We have therefore shown that FCQEM using Equation~\ref{eq:FCQEM_full} can find exact eigenvalues and performs similarly to VD for HeH$^+$ not only with the exact eigenstate but also trial states that are rotated from the eigenstate. 

In general, however, the effectiveness of FCQEM depends on how the trial state appears in the Pauli basis that dominates the decomposition in Equation~\ref{eq:paulidecomp}. From Figure~\ref{fig:H2}(c), we can see there exist states where FCQEM does worse than VD and, from Equation~\ref{eq:err_deriv}, there may be directions where FCQEM does not maintain the eigenvalue on perturbation. In particular, better error mitigation will occur when the distribution has sharp peaks (equivalent to VD for exact eigenstates), whereas FCQEM has no effect for uniform distributions.
This implies that practitioners using quantum algorithms such as VQE to find ground state with FCQEM may need to restrict their Ansätze to states that minimise the FCQEM truncation error, which may not necessarily be the exact ground state.
For arbitrary state preparation algorithms or Ansätze that are not sufficiently expressive, we can combine our method with the Quantum Computed Moments (QCM) method~\cite{hv2020}, which is able to get ground state energies and arbitrary ground state observables from such inexact trial states. Furthermore, performing FCQEM on expectation values can remove significant noise before input into QCM. In particular, it requires no quantum overhead and can be performed with classical post-processing for QCM. We now detail how to combine both FCQEM and QCM to improve on each individual method.

\subsection{Combining with Quantum Computed Moments Method}

The QCM method~\cite{hv2020} has emerged as a powerful quantum heuristic for estimating the ground state energy of spin models and molecular Hamiltonians~\cite{Jones2022, Jones2024}. In this approach, expectation values of powers of the Hamiltonian (or \emph{moments}) $\{\langle H \rangle, \langle H^2 \rangle, \langle H^3 \rangle, \langle H^4 \rangle\ldots\}$ are computed with respect to an approximate trial state with some overlap with the ground state (e.g. a variational Ansatz with suboptimal parameters), and combined using a cluster expansion of the Lanczos iteration~\cite{analytic} to produce an estimate of the true ground state energy $E_0$. To fourth order in the moments, the expression takes the following form~\cite{el4}:
\begin{equation}
E_0 \approx c_1 - \frac{c_2^2}{c_3^2 - c_2 c_4} \left(\sqrt{3 c_3^2 - 2 c_2 c_4} - c_3\right),\label{eq:qcm}
\end{equation}
where the $c_n$ are cumulants given by:
\begin{equation}
c_n = \langle H^n\rangle - \sum_{p=0}^{n-2}\binom{n-1}{p}c_{p+1} \langle H^{n-1-p}\rangle.
\end{equation}
The QCM ground state energy estimate has been shown to greatly improve on the direct energy expectation value $\langle H \rangle$ (the first moment) for trial states prepared on real quantum hardware in the presence of noise~\cite{hv2023}. While we will only focus on the ground state energy here, the QCM approach may be extended to estimate arbitrary observables with respect to the ground state, using only approximate noisy trial states~\cite{hv2023_2, Jones2025}.

In practice, the moments $\{\langle H \rangle,\langle H^2 \rangle,\langle H^3 \rangle,\langle H^4 \rangle\}$ required for the ground state energy estimate in Equation~\ref{eq:qcm} may be computed as follows:
\begin{enumerate}
    \item Express the Hamiltonian $H$ as a weighted sum of Pauli strings, $H = \sum_iw_iP_i$ and multiply term-by-term to obtain expressions for $H^2$, $H^3$ and $H^4$ as weighted sums of Pauli strings. Find $\mathcal{P}$, the set of Pauli strings that appear in one or more of these expressions.
    \item Determine a set $\mathcal{Q}$ of Pauli tensor product basis (TPB) measurements needed to find the expectation value of every Pauli string ($\mathcal{Q}$ is typically much smaller than $\mathcal{P}$ since many $P\in\mathcal{P}$ will qubit-wise commute).
    \item Prepare the trial state $\ket{\phi}$ on a quantum computer, followed by the relevant basis transformation (one layer of single-qubit gates) to measure $\ket{\phi}$ ($n_{\mathrm{shots}}$ times) in each TPB $Q\in\mathcal{Q}$.
    \item Post-process output to determine Pauli expectation values $\langle \phi | P |\phi\rangle$ for each $P\in\mathcal{P}$, which are then used in the original expressions to find moments $\{\langle H \rangle,\langle H^2 \rangle,\langle H^3 \rangle,\langle H^4 \rangle\}$.
\end{enumerate}

While this procedure works well in most cases, noise on the individual moments can induce specific pathological behaviour. Observing Equation~\ref{eq:qcm}, if the square root is negative, or if the denominator is zero, then the moments corrected energy becomes unphysical. We want to correct the moments with FCQEM before using Equation~\ref{eq:qcm} to minimise errors in the final ground state energy estimate.

FCQEM is particularly suited to correcting moments of the Hamiltonian since they all commute, i.e. $[H^k,H] = 0,~\forall k\in\mathbb{N}$, so truncation errors remain the same. In addition, there is no quantum overhead, so we can run QCM normally and use FCQEM as classical postprocessing to either square each probability for each TPB as per Equation~\ref{eq:FCQEM_full} or combine two separate sets of sampling statistics as per Equation~\ref{eq:FCQEM2} for each moment.

We will now demonstrate moments method calculations on a real superconducting device, using the same output data with FCQEM post-processing to see how mitigating the noise on each moment can improve the QCM ground state energy estimate with respect to an inexact noisy trial state.

\subsection{Experimental Results on Rigetti Quantum Processor}

\begin{figure*}[t!]
    \centering
    \includegraphics[width=\linewidth]{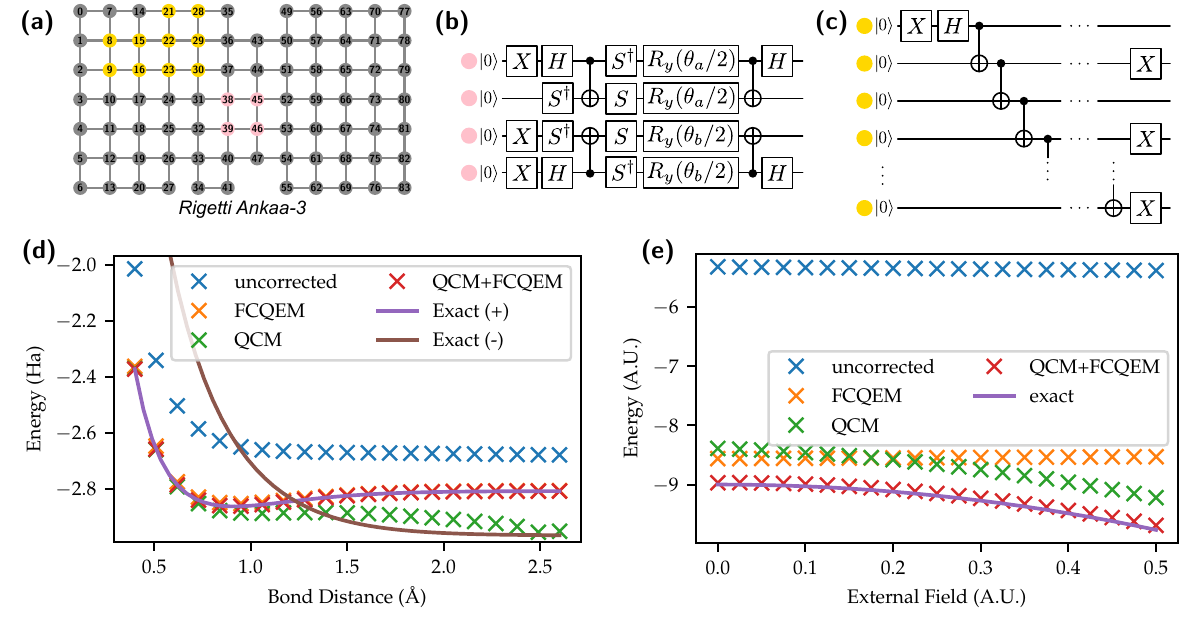}
    \caption{\justifying \textbf{Implementation on Rigetti Quantum Processors.} Experimental demonstration of FCQEM and QCM on Rigetti Ankaa-3 84-qubit superconducting device for Helium Hydride (HeH$^+$) and 10 site Transverse Field Ising Model (TFIM) Hamiltonians. (a) The qubit layout for Ankaa-3 is given with pink (yellow) colour to indicate qubits used to prepare the trial state for the HeH$^+$ (TFIM) and yellow colour to indicate qubits for TFIM. (b) The UCCS trial state circuit and (c) N\'eel state circuit. (d) The HeH$^+$ results as the bond distance between He and H atoms increases. FCQEM is able to correct most of the noise but does not recover the exact ground state energies. At short bond lengths, QCM and FCQEM+QCM can recover the energies and, since FCQEM requires no quantum resources, shows they are efficient to combine. At long bond lengths, QCM finds the negative charge state as it is a lower energy state, whereas QCM+FCQEM recovers the correct charge state. Similarly, (e) shows the TFIM results. For no field, FCQEM outperforms QCM as this state is optimal and the Hamiltonian is diagonal, however FCQEM+QCM is required to recover the ground state energy. As the field increases, FCQEM cannot correct past the N\'eel state, whereas QCM can, and QCM+FCQEM is able to recover the ground state energy for small external fields with worse performance as the Hamiltonian becomes less diagonally dominant.}
    \label{fig:noise}
\end{figure*}

To demonstrate the efficacy of our FCQEM method in conjunction with the QCM technique, we perform two sets of experiments compiled~\cite{smith2016} on 82-qubit Rigetti Ankaa-3 superconducting quantum processor~\cite{Karalekas2020}. The device noise resulted in a median iSWAP fidelity of $98\%$ and a median readout fidelity of $97\%$. (a) shows the schematic diagram of quantum processor layout, along with the selection of qubits for the two experiments indicated by the yellow and red colors. In the first experiment, we implemented the 4-qubit HeH$^+$ Hamiltonian defined in the previous section on qubits 38, 39, 45, and 46. A UCCS Ansatz was prepared and optimised for $1$~Å. It was unable to find the exact ground state since it does not include double excitations and serves as an inexact trial state. In the second experiment, 10-qubit N\'eel state circuit was compiled on the Rigetti processor qubits 8,9, 15, 16, 21, 22, 23, 28, 29, and 30 to compute the ground state energy of 1-D transverse-field Ising model (TFIM). These two experiments were carefully chosen to demonstrate complimentary advantages of the FCQEM and QCM methods, as well as their combined accuracy.  
\\ \\
\noindent
\textbf{HeH$^+$ Molecule:} Figure~\ref{fig:noise}. (b) plots the HeH$^+$ trial state circuit whose mapping on the Rigetti processor is indicated by the red color qubits in (a). We prepared the HeH$^+$ UCCS ground state for $1$~Å bond length in (d) and measured the expectation values for Hamiltonians at varying bond lengths between the Helium and Hydrogen atoms. The results from the device were always significantly higher than the ground state energy due to noise (blue crosses). Applying FCQEM by itself was able to correct most of the noise, however it was not able to recover the exact ground state energy (orange crosses). In comparison, for short bond lengths, QCM was able to find the exact ground state energies, but significantly deviated from the exact energy for larger bond lengths (green crosses). 

At long bond lengths, QCM goes below the exact energy for the positive charge state. This is because QCM will find the ground state of the exact matrix it is given and ignore restrictions set by the Ansatz. We are using a second quantised Hamiltonian with a Jordan-Wigner mapping which means the qubit Hilbert space contains all charge states up to full occupation of the STO-3G bases.  We performed Hartree-Fock (HF) to find the Hartree-Fock state and molecular orbital basis for the positive charge state, which means that restricting to the HF state and charge preserving excitations will fix the charge state. In this work, we use a UCCS Ansatz, which has limited expressivity to only the Hartree-Fock state and single excitations that preserve the number of electrons. This could also be done by Hamiltonian evolution or other guiding state methods. The problem is that the noise, in particular depolarising and readout error, adds contributions from states in the other charge subspaces. Since QCM corrects to the lowest energy eigenvector given sufficient overlap, the noise adds enough overlap that QCM corrects to the lower energy charge states. However, if we apply FCQEM first, we can sharpen the distribution towards the HF state and remove small contributions from other states due to noise. We can therefore see that applying FCQEM before QCM can recover the positive charge state energy for all bond lengths as it reduces the overlap with the ground states in other charge subspaces or even the full space.

\begin{figure}
    \centering
    \flushleft{(a)}
    \includegraphics[width=\columnwidth]{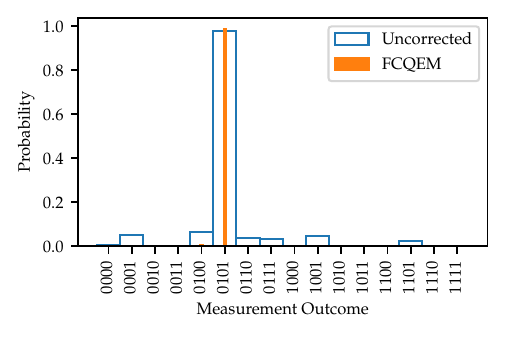}
    \flushleft{(b)}
    \includegraphics[width=\columnwidth]{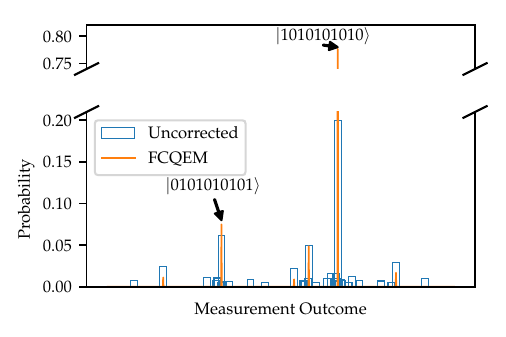}
    \caption{\justifying \textbf{Probability distributions from Rigetti Experiments.} Probability distributions from computational measurement of Figure~\ref{fig:noise}. Distirbutions are shown uncorrected from device and squared with normalisation to implement FCQEM. FCQEM amplifies peaks and suppresses small contributions for each POVM. (a) shows HeH$^+$ with UCCS Ansatz distribution; (b) shows transverse field ising model with N\'eel state distribution. Blue boxes represent the uncorrected values with orange lines at the FCQEM corrected values. Values below a probability of $0.005$ are omitted for clarity.}
    \label{fig:probdist}
\end{figure}

In summary, our experiment confirm that combining FCQEM with QCM estimates the ground state energy for the entire spectrum of bond lengths within $10^{-5}$ Ha accuracy which is better than the chemical accuracy requirement. We also note that since FCQEM was implemented by squaring the probability distributions, as per Figure~\ref{fig:probdist}, it can be performed with no extra quantum resources and is therefore not detrimental to combine with QCM even when QCM recovers the ground state energy such as at the low values of bond lengths in this experiment. 
\\ \\
\noindent 
\textbf{Transverse-field Ising Model}: As it is a common benchmark for quantum error mitigation techniques~\cite{Kim2023}, we also apply our method to estimate the ground state energy of the transverse-field Ising model (TFIM):
\begin{equation}
    H = \sum_{\langle i\,j\rangle}J_{ij}Z_iZ_j + h\sum_iX_i,
\end{equation}
where the $\{\langle i\, j\rangle\}$ are adjacent sites on a lattice, $J_{ij}$ is the corresponding coupling strength, and $h$ is the external field strength. We choose an $n$-qubit antiferromagnetic N\'eel state:
\begin{equation}
    |\phi^-\rangle=\frac{1}{\sqrt{2}}\left(|01\rangle^{\otimes \frac{n}{2}}-|10\rangle^{\otimes \frac{n}{2}}\right),
\end{equation}
which may be prepared via the circuit shown in Figure~\ref{fig:noise}(c).

This trial state is an exact ground state at $h=0$ but not a good approximation of the ground state at greater values of $h$, since its energy does not change (the transverse field term always has zero expectation value). While we thus cannot expect FCQEM alone to recover the ground state energy for $h>0$ from this trial state, its overlap with the true ground state permits QCM to correct towards the true ground state energy. Combining the two methods should allow us to study the ground state energy of this model with respect to $h$ by only needing to prepare the one trial state.

The 10-qubit N\'eel state circuit was compiled~\cite{smith2016} and run on the Rigetti Ankaa-3 superconducting quantum processor~\cite{Karalekas2020}, using the yellow color qubits shown in Figure~\ref{fig:noise}(a). Figure~\ref{fig:noise}(e) shows the ground state energy estimates of the 10-qubit 1D TFIM as a function of field strength, $h$, using the antiferromagnetic N\'eel state $\ket{\phi^-}$ as a trial state on the Rigetti device. FCQEM, QCM and the combination of both methods are used to correct the energy obtained from the noisy device towards the true ground state energy. When $h=0$, FCQEM offers a better correction than QCM for the fully diagonal Ising Hamiltonian. As $h$ increases, and the Hamiltonian is more dominated by Pauli terms corresponding to measurement bases that express $\ket{\phi^-}$ as a uniform distribution, QCM is more effective than FCQEM at estimating the ground state energy. However, in both regimes, the combined QCM+FCQEM approach consistently improves on both methods, recovering the true ground state energy to high precision. We therefore emphasise that it is often worth implementing FCQEM when using QCM even if it offers minor improvement on the ground state energy estimate, due to the negligible additional computational cost.

To visualise the mechanism of FCQEM, we plot the probability of measuring each outcome in the computational basis for the experimental results in Figure~\ref{fig:probdist}. The corrected case is given by:
\begin{equation}
    \frac{p_{Z,i}^2}{\sum p_{Z,i}^2},
\end{equation}
for each outcome $i$. It can be seen that the corrected distribution sharpens the peaks of the noise-free distribution and suppresses small non-zero amplitudes which are due to noise. For Figure~\ref{fig:probdist} (a), this corresponds to increasing the Hartree-Fock state contribution, which is the computational basis state with the largest contribution. In particular, as mentioned from Figure~\ref{fig:noise}, this allows us to recover the correct charge state for the Hartree-Fock state.

Similarly, for Figure~\ref{fig:probdist} (b), we show the N\'eel state which should have two equal peaks at $\ket{1010...10}$ and $\ket{0101...01}$. Squaring the distribution with FCQEM amplifies these two peaks and reduces the contribution from the smaller peaks introduced by noise. FCQEM also amplifies the noise-induced asymmetry between the $\ket{1010...10}$ and $\ket{0101...01}$ amplitudes, which in this specific case does not inhibit the correction toward the ground state energy, as both of these states are ground states of the $h=0$ antiferromagnetic Ising model. This supports why FCQEM alone could outperform QCM in noise mitigation for small $h$ in Figure~\ref{fig:noise}(e). We note that the combination of both techniques was still significantly more effective than either alone (with an approximation error between $0.3\%$ and $0.8\%$ for $h<0.5$), which is especially useful as it requires negligible computational overhead to incorporate FCQEM into QCM. 

\begin{figure*}
    \centering
    \includegraphics[width=\textwidth]{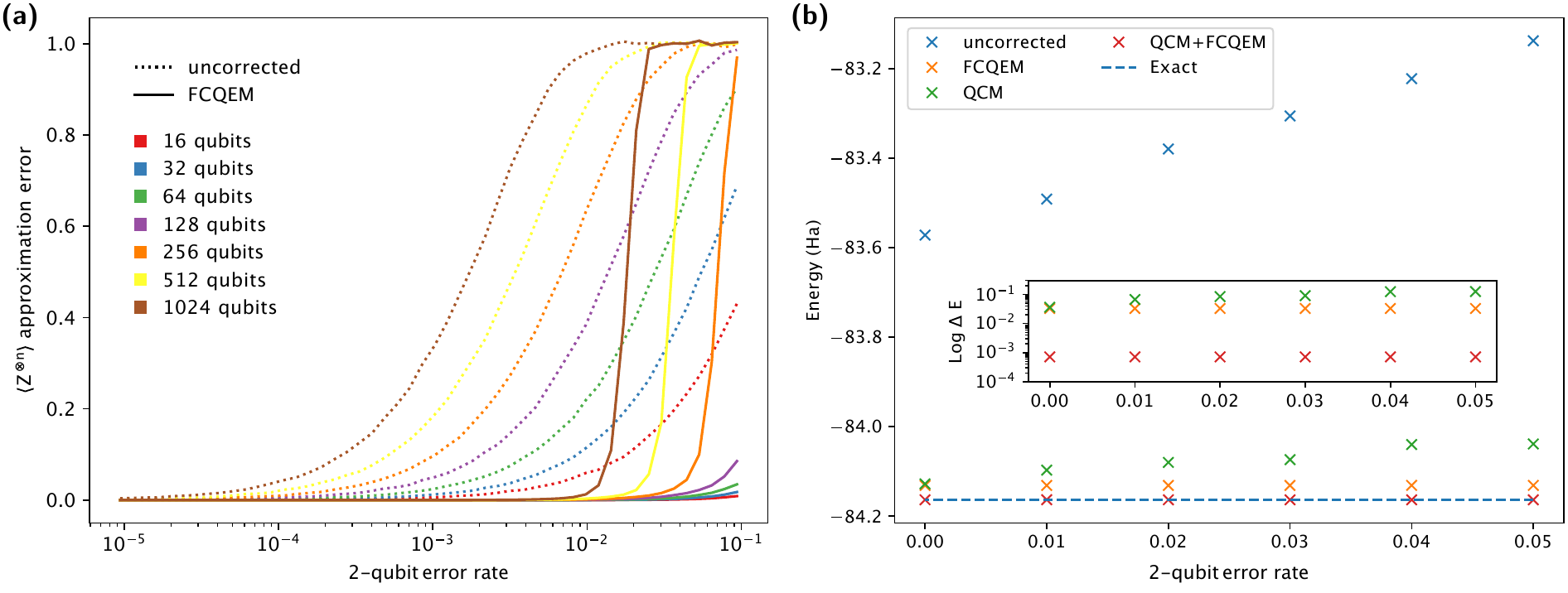}
    \caption{\justifying \textbf{FCQEM Scalability.} Demonstration of scaling of FCQEM with larger simulations. (a) shows various scales (16-1024 qubits) of the effectiveness of FCQEM at determining a $n$-qubit spin correlation observable $\langle Z^{\otimes n} \rangle$ with respect to the antiferromagnetic N\'eel state, prepared with noise using a stabiliser simulator. The raw expectation value (dotted line) and FCQEM correction (solid line) are shown as a function of error rate, where we have applied a Pauli noise channel biased toward dephasing (as to mimic NISQ device noise at the $10^{-2}$ level). Each simulation is run with 100\,000 shots. (b) shows the 8 qubit H$_2$O molecule Hamiltonian and Ansatz from~\cite{Jones2024} under varying depolarisation noise. The noise is given as two qubit depolarisation on each CNOT gate with a tenth of the depolarisation on each single qubit rotation. Inset of (b) shows the difference from exact ground state energy on a logarithmic scale. The circuit from~\cite{Jones2024} does not prepare an exact ground state but QCM was able to recover the ground state energy. Here we show that FCQEM is not only able to find energies with a slightly better accuracy compared to QCM but noticeably the combination of the two methods increases accuracy by two orders of magnitude and is less sensitive to depolarisation noise. }
    \label{fig:scaling}
\end{figure*}

\subsection{Scalability}
Since the circuit to prepare the N\'eel state is Clifford, a stabiliser simulation was also employed~\cite{gidney2021stim} to study the effectiveness of FCQEM against Pauli noise at scales up to 1024 qubits. The Pauli error channel was constructed as to mimic the error profile seen on NISQ devices, where dephasing has an order of magnitude greater effect than depolarising noise.

Since pre-processing for QCM is computationally expensive at such scales, we shift our focus from ground state problems (where FCQEM is applied to a number of Hamiltonian moments) to using FCQEM alone to mitigate noise in the expectation value of a single observable. Figure~\ref{fig:scaling} shows the approximation error of FCQEM in computing the $n$-qubit spin correlation observable $\langle Z^{\otimes n} \rangle$ with respect to the antiferromagnetic N\'eel state $\ket{\psi^-}$, simulated at varying noise levels and numbers of qubits. This demonstrates at scale the robustness of FCQEM when truncation error is small. When two-qubit error rates are at $10^{-2}$ (approximately NISQ-level) and the number of shots is fixed at $100\,000$, FCQEM is able to correct the observable to its noise-free expectation value for instances up to 1024 qubits. The error mitigation attainable by FCQEM only breaks down when a shot-noise limit is reached, beyond which noisy distributions look mostly uniform.

We also simulate a larger molecule (H$_2$O, the water molecule) to demonstrate scalability of FCQEM with QCM. We use the 8-qubit Hamiltonian and Ansatz circuit from Jones et al.~\cite{Jones2024}, specifically the circuit that only uses four excitations and cannot prepare the exact eigenstate. We simulated the circuit with $10,000$ shots with one and two-qubit error rates. Two-qubit error rates were fixed at ten times the one-qubit rate and the two-qubit rate was set at a NISQ-level order of magnitude from $0$ to $0.05$. Figure~\ref{fig:scaling} (b) shows the results in a linear scale with an inset for logarithmic scale for the corrected measurements. FCQEM performs identically to QCM for no noise and corrects an order of magnitude closer to the exact energy. Combining them improves this correction by two more orders of magnitude. QCM is more sensitive to increasing depolarisation than FCQEM. When we combine them, since we are taking the FCQEM distributions and applying QCM to it, it is able to retain the insensitivty to noise and recover the exact energy to within $10^{-3}$ Ha, reaching the chemical accuracy limit. 

\section{Conclusion}
In this work we have introduced Fictitious Copy Quantum Error Mitigation (FCQEM), a purely classical post-processing technique that mitigates noise in expectation values by exploiting powers of measured probability distributions. We showed that FCQEM is formally equivalent to a first-order truncation of Virtual Distillation and that, for eigenstates of an observable, it reproduces the exact noise-free expectation values without requiring additional quantum resources, entangling operations, or the preparation of multiple noisy copies of a state.
We analysed the conditions under which FCQEM is effective and demonstrated that its performance depends on the structure of the measurement basis and the resulting probability distributions. In particular, FCQEM performs optimally for sharply peaked distributions and has no effect for uniform distributions. This naturally restricts its general applicability to eigenvalue problems, requiring Hamiltonians that are dominated by a single Pauli basis and trial states which permit sharply peaked distributions, however this a class that nevertheless includes many physically relevant models such as configuration interaction Hamiltonians in quantum chemistry and Ising-type spin models.
Crucially, FCQEM is compatible with other error mitigation and post-processing strategies such as the Quantum Computed Moments (QCM) method, which allows us to move past the trial state limitation to find the true ground state energy. FCQEM improves the robustness of QCM ground state energy estimation in noisy regimes and the combined approach mitigates state-dependent truncation errors. Across both simulated noise models and real experiments on Rigetti superconducting devices, we showed that FCQEM alone can substantially reduce noise in expectation values, and that the combined FCQEM+QCM approach consistently improves ground state energy estimates relative to either method used independently.
Beyond quantum systems, we also demonstrated that the underlying mechanism of FCQEM applies to entirely classical probability distributions, highlighting that squaring and renormalising distributions generically sharpens peaked structure and improves expectation value estimates under broad classes of noise. This perspective clarifies the classical origin of the mitigation effect and provides intuition for when FCQEM can be expected to succeed or fail.
Overall, FCQEM offers a simple, low-cost, and scalable noise mitigation tool for near-term quantum devices. Its negligible computational overhead and ease of integration make it particularly attractive as a default post-processing step, especially in conjunction with methods such as quantum computed moments (QCM). More broadly, our results suggest that a careful separation of genuinely quantum advantages from classical post-processing effects is essential when assessing the power of near-term quantum error mitigation techniques.

\section{Acknowledgements}
This research was supported by Amazon Web Services (AWS) by providing access to quantum computing resources via Amazon Braket that enabled experimental demonstrations in this work.

\bibliography{refs}
\end{document}